# *Surface impedance measurements in superconductors in dc magnetic fields: challenges and relevance to particle physics experiments.*

Andrea Alimenti, Nicola Pompeo, Kostiantyn Torokhtii and Enrico Silva

Particle physics and radio-frequency (rf) superconductivity have driven each other on since the 1970s. The unique properties of superconductors (SC) have been the enabling keys for the realization of accelerators with always increased performances thanks to the realization of all-superconducting cavities. The use of increasingly pure superconducting coatings for accelerating cavities, with lower and lower rf losses, determined such high quality factors [1] that the need to operate at low temperatures (below the superconducting transition temperature $T_c$) was well paid for. Recently, with respect to the path followed by the high frequency superconductivity [2], a new field opened since SCs are being considered for GHz operation in high dc magnetic fields, and measurements (and optimization) of totally different quantities are needed. The possibility of successfully using SCs in high magnetic fields for these purposes is far from obvious and it depends on the outcome of accurate measurements of usually overlooked quantities.

In fact, new accelerators such as the Future Circular Collider (FCC) at CERN [3] or the Super Proton–Proton Collider SPPC in China [4] will need 100-km long beam screens able to screen the synchrotron radiation with power spectra up to ~ 2 GHz, in dc fields up to 16 T, and at temperatures $T \sim 50$ K. Low surface impedance $Z$ materials are needed to this aim, with $Z = \frac{E_\parallel}{H_\parallel} = R + \mathrm{i}X$, $E_\parallel$ and $H_\parallel$ are the electric and magnetic field components parallel to the surface of the materials, $R$ is the surface resistance and $X$ is the surface reactance. It is still unclear whether copper is a suitable solution, hence high-$T_c$ SCs are under scrutiny to assess if they can be once more the enabling solution.

In parallel, in the context of the hunt for dark matter a new generation of hybrid super/normal-conductive resonating cavities operating at several GHz, known as haloscopes, is being developed to detect the radio-frequency/microwave photon in which a virtual photon is converted by a supposed dark-matter constituent, an axion, in presence of magnetic fields. The sensitivity of these detectors is $\propto Q_u B^2$ [5], with $Q_u$ the cavity quality factor and $B$ the externally applied magnetic flux density. SCs are then an obvious possible solution.

Finally, and looking further ahead in time, a future generation of muons circular colliders would be based on muon cooling radio frequency (0.5-1 GHz) cavities, which will operate in stray fields up to 5 T. Again, materials with low $Z$ in high magnetic fields are a key enabler and SCs are taken into consideration [6].

The new player to be challenged in the rf superconductivity is the dc magnetic field: all technologically relevant superconductors are so-called type-II superconductors. There, in even moderate dc fields $H > H_{c1}$, with the lower critical field $\mu_0 H_{c1}$ of the order of $(10^1 - 10^2)$ mT depending on the material, magnetic flux tubes known as 'fluxons' proliferate, driving the superconductor in the so-called 'mixed state'. Quantum mechanics dictates that each fluxon carries exactly a magnetic flux quantum $\Phi_0 = \frac{h}{2e} \approx 2.067 \times 10^{-15}$ Wb ($h$ is the Planck constant, $e$ is the electric charge quantum), sustained by lossless circular superconducting currents (whence also the name of 'vortices'). The challenge resides in the fact that external currents exert an effective Lorentz force on fluxons, which can set them in motion. Due to the normal core of the fluxons and the related scattering phenomena, the motion of the fluxons is highly dissipative. Thus, pinning the fluxons on 'pinning centres' (PCs), which are defects of the material lattice, is a central issue in material science [7]. Rf and microwave currents are even trickier, since they produce an alternating fluxon motion that is more difficult to hinder.

The frequency dependence of the radio-frequency/microwave power losses in superconductors in dc magnetic fields was first shown in a seminal experimental study in 1966 [8], as reported in Figure 1(a). Two important features of the real part of the material resistivity Re($\rho$) (proportional to the material power absorption $P$ in electromagnetically thin samples) emerge. (i) The presence of a characteristic frequency $f_c$, which is proportional to the strength of the pinning force acting on fluxons and that marks the crossover between the low-frequency, low dissipation regime and the high-frequency, high dissipation regime. Thus, information on $f_c$ is essential to engineer suitable SC materials for rf applications: as a rule, for applications at a given frequency $f$ one looks for $f \ll f_c$. (ii) The high frequency saturation value to which Re($\rho$) approaches, called flux-flow resistivity $\rho_{ff}$, corresponding to the effective resistivity of completely free moving fluxons. The flux flow resistivity $\rho_{ff}$ is of the order of the fraction $H/H_{c2}$ of the resistivity in the normal state $\rho_n$, $\rho_{ff} \sim \rho_n H/H_{c2}$ [9] with $H_{c2}$ the upper critical field above which the superconductivity is suppressed. Thus, $\rho_{ff}$ can attain rather large values.

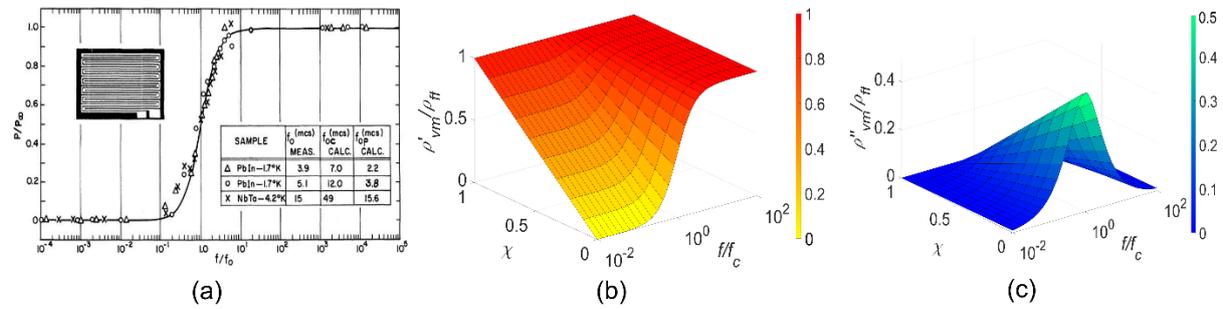

*Figure 1 – Vortex motion resistivity in superconductors as a function of the normalized frequency $f/f_c$: (a) original experimental data for the power absorption in low-$T_c$ superconductor [8] (here $f_0 = f_c$). https://doi.org/10.1103/PhysRevLett.16.734 (b) Calculated normalized real resistivity and (c) calculated normalized imaginary resistivity with the inclusion of the thermal effects.*

Since 1966 many other, more technologically suitable SCs were discovered, with higher $T_c$ and $H_{c2}$. Nb$_3$Sn ($T_c \approx 18$ K) has been employed since long in high-field magnets, but its potential for rf superconductivity is still unexpressed. Obvious interest is also in High $T_c$ Superconductors (HTS), with transition temperatures $T_c \approx 90$ K for, e.g., YBa$_2$Cu$_3$O$_{7-x}$, (YBCO). However, high operating temperatures imply a relevant role of thermally activated fluxon jumping between equilibrium sites, resulting in so-called flux creep dissipation. In high-frequency superconductivity, flux creep is measured by the adimensional normalized parameter $0 \leq \chi \leq 1$ which weighs the effect of thermally excited fluxon jumps out of the pinning well, from zero jumps ($\chi = 0$) to jumps so frequent ($\chi = 1$) that the pinning wells effect is practically irrelevant.

The three parameters $f_c$, $\rho_{ff}$ and $\chi$ are the main quantities of interest and require a nontrivial combination of methods and models to be determined. A metrological approach involves (i) the introduction of an appropriate physical model relating the parameters to directly measurable quantities, (ii) a suitable experimental setup, with a study of the uncertainties and of the methods to reduce them, and (iii) experimental results for the desired quantities. These aspects are presented in the following sections, thus providing a focus on the measurement methods, and on their main critical issues, useful for the characterization of SCs in the operating conditions of the aforementioned applications. Short conclusions with a view on unresolved aspects and future developments are reported in the end. It should be stressed that, although the path here depicted has been explored in moderate fields in the past [8, 10], the need for reliable measurements in high magnetic fields is somewhat a game changer due to the increased complexities.

## Physical background

A more complete physical picture beyond Figure 1(a) is essential to define the measurement strategy and to understand and support the technological requirements.

First, it is important to note that the zero-field contribution to the surface impedance, so relevant for accelerating cavities, is negligible with respect to the fluxon motion contribution for even small magnetic fields ~ 0.1 T. Thus, in the *linear response* (i.e. $E \propto J_{rf}$ with $E$ the electric field and $J_{rf}$ the radio frequency current densities inside the conductor), the "material property" ac resistivity $\rho$ essentially coincides with the vortex motion resistivity $\rho_{vm}$ which, in turn, on very general grounds, can be described through the following [11]:

$$\rho_{vm} = \rho_{ff} \frac{\chi + i\frac{f}{f_c}}{1 + i\frac{f}{f_c}} = \rho'_{vm} + i\rho''_{vm}. \tag{1}$$

Eq. (1) ceases to apply when the fluxon oscillations around their equilibrium positions (PCs, typically) is a sizeable fraction of their separation, as it is expected at low frequencies (below ~ 1 GHz) or at high rf drive, so that non-linear regimes require a specific theoretical and experimental treatment.

The vortex parameters $f_c$, $\chi$ and $\rho_{ff}$ are each connected to a specific physical mechanism, as depicted in the introduction. We refer to specialized reviews [10, 12] for an in-depth discussion on their nature and roles.

With $\chi = 0$ Eq.(1) reproduces the data taken on low-$T_c$ superconductors of Fig. 1(a), and it has been for long time the reference model for rf vortex motion [8, 10, 13]. The full model however foresees a frequency dependence where the role of $\chi$ can be relevant, as shown in Figure 1(b)-(c). As it can be seen, at a given frequency $f$, $\chi$ and $f_c$ determine the fraction of the asymptotic dissipation $\rho_{ff}$ and thus the overall response of the superconductor.

Wideband measurements (over more than a decade in frequency around $f_c$) with the so-called Corbino disk [14, 15] validated the model on several SCs and assessed its range of applicability showing that below a few GHz a more complex description applies [14]. Nevertheless, the reliable determination of $\chi$, $f_c$ and $\rho_{ff}$ is the main metrological challenge in high-frequency measurements on SCs in dc magnetic fields.

## Materials and methods

The surface impedance $Z = R + iX$ of several SC samples is measured in dc magnetic fields to obtain $\rho_{vm}$ and the vortex motion parameters described in the previous section. In this paper we show the results obtained on the materials and samples detailed in Table 1, for the purpose of comparing their performances (see Fig. 5). The functional relation $Z(\rho_{vm})$ depends on the geometry of the sample and it will be detailed in the experimental section in the different case studies presented.

*Table 1 - Samples list*

| *Ref.* | *Material* | *Sample type* | *Substrate* | $T_c$ *(K)* | *Perspective high-frequency applications in particle physics* |
|---|---|---|---|---|---|
| S1 [16] | YBa$_2$Cu$_3$O$_{7-x}$ | 80 nm thick film | LaAlO$_3$ | ~ 91 | beam screen coating, haloscopes, |
| S2 [17] | YBa$_2$Cu$_3$O$_{7-x}$ +5 % BaZrO$_3$ | 100 nm thick film | SrTiO$_3$ | ~ 91 | |
| S3 [18] | Nb$_3$Sn | bulk | __ | ~ 18 | haloscopes, RF cavities |
| S4 [19] | FeSe$_{0.5}$Te$_{0.5}$ | 240 nm thick film | CaF$_2$ | ~ 18 | still unexplored |
| S5 [20] | MgB$_2$ | bulk | __ | ~ 39 | RF cavities |

Even if wideband measurements spanning more than a decade in frequency would be the most direct way to determine the vortex motion parameters, such methods have intrinsically low sensitivity, being usually relegated to the measurements on medium to high loss materials. In fact, wideband methods are used only in the very high dissipation regime [21, 22]. When high sensitivity is needed, resonant measurement methods are preferred [23]. In fact, resonant methods can even be used to study the residual surface resistance of SCs, which can be of the order of $10^{-9}$ $\Omega$ at few GHz [1], while the resolution of traditional wideband reflection methods are usually limited at some tens of m$\Omega$ [14, 21].

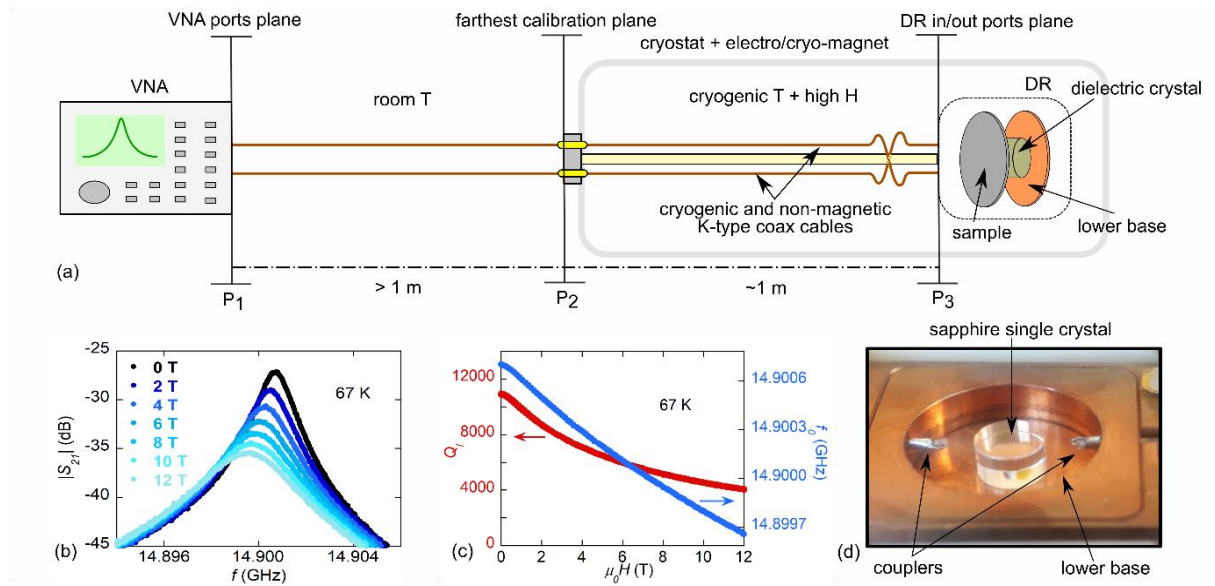

*Figure 2 - (a) Sketch of the measurement line. The Vector Network Analyzer (VNA) is linked to the DR through a >1 m long room temperature line (between $P_1$ and $P_2$) and a ~ 1 m long cryogenic and non-magnetic coaxial line (between $P_2$ and $P_3$). This last part of the microwave line is made in phosphor-bronze to withstand the cryogenic and magnetic environment. (b) Transmission S-parameter resonance curves measured on a $YBa_2Cu_3O_{7-x}$ sample, keeping the temperature within $(67 \pm 0.05)$ K, at selected magnetic field magnitudes between 0 T (black) and 12 T (light blue). (c) Loaded quality factor $Q_l$ and resonance frequency $f_0$ measured through the Lorentzian fit (model shown in Eq.(6)) of the $S_{21}(f)$ curves shown in (b). (d) Bitonal copper-shielded DR, loaded with a cylindrical sapphire single crystal (diameter $7.13(1)$ mm, height $4.50(1)$ mm).*

A powerful resonant method to measure the surface impedance in superconducting samples relies on the use of dielectric loaded resonators (DRs) [13, 24-27]. A DR is a metallic cavity loaded with a low loss dielectric crystal in order to increase the measurement sensitivity. In Figure 2 a sketch of a typical measurement system, a picture of a DR used at Roma Tre University, and some measured resonance curves are shown. The sample under investigation is loaded as an end-wall of the DR and its $Z$ is obtained from the resonator unloaded quality factor $Q_u$ and resonance frequency $f_0$ through the following equations [23, 25]:

$$\frac{1}{Q_u} = \frac{R}{G} + bckg_Q , \qquad (2)$$

$$\frac{\Delta f_0}{f_{0,ref}} = -\frac{\Delta X}{2G} + bckg_{f_0} , \qquad (3)$$

$G$ is the geometrical factor of the sample loaded into the DR at the electromagnetic resonance mode of interest and $\Delta y = y - y_{ref}$ indicates the variation of the quantity $y$ with respect to the reference value $y_{ref}$. Finally, $bckg_Q$ and $bckg_{f_0}$ are the background contributions on $Q_u$ and $f_0$, respectively, given by the resonator itself.

Relevant metrological aspects, with the contextual identification of the main sources of uncertainties, can be derived from Eq.s (2) and (3):

1. The quantities of interest ($R$ and $X$) can be obtained after the $Q_u$ and $f_0$ measurement, only if $bckg_Q$ and $bckg_{f_0}$ are removed through a calibration of the DR. The DR calibration would require a conducting standard, of known $R_m$ and $X_m$, to be loaded in place of the sample, so that, by inverting Eq.s (2) and (3), $bckg_Q$ and $bckg_{f_0}$ could be derived from $Q_u$ and $f_0$ measurements. However, no conductive standards exist with $R_m$ and $X_m$ known with high enough accuracy, in the whole magnetic field, cryogenic temperature, and microwave frequency ranges desired. A differential end-wall replacement perturbation method [23] is preferred instead: the variations on $Q_u$ and $f_0$ with respect to a reference state of the sample are used to obtain the variations $\Delta Z$ considering, in the small perturbation limit, that all the other instrumental contributions are not changed. In particular:

$$\frac{1}{Q_u} - \frac{1}{Q_{u,ref}} = \frac{R - R_{ref}}{G}, \qquad (4)$$

$$\frac{f_0 - f_{0,ref}}{f_{0,ref}} = -\frac{X - X_{ref}}{2G}. \qquad (5)$$

The subscript '$ref$' is referred to the reference, and with this approach the problem of the DR calibration is overcome. Since $bckg_Q$ and $bckg_{f_0}$ depend on the temperature $T$, this technique is particularly accurate for measurements at fixed $T$ and varying magnetic field, provided that the DR is specifically realized with non-magnetic materials. Thus, at fixed $T$ Eq.s (4) and (5) directly yield the variations on $R$ and $X$ using the zero field state as the reference state, and the delicate and uncertainty-source calibration of the DR can be avoided altogether [25]. Obviously, this approach completely rules out the use of superconducting cavities instead of (normal metal-based) DR to exploit their higher $Q$ and thus higher sensitivities, since their magnetic-dependent response would yield a heavily field dependent background not discriminable from the sample magnetic dependent $Z(H)$.

An estimation of the uncertainty linked to imperfect thermalization during the measurements, which represents the main uncertainty source within this differential approach, can be provided as follows: the DR without a SC sample, at zero field, has variations $\Delta Q \sim 300$ and $\Delta f_0 \sim 100$ kHz when the temperature changes from 5 K to 25 K. With a thermalization within ± 0.05 K, the standard for the measurements here presented, one can give a rough estimate $\Delta Q \sim 0.75$ and $\Delta f_0 \sim 250$ Hz, fully negligible with respect to the measured variations due to the SC response (see Fig. 3).

2. The geometrical factor $G$ must be accurately determined. However, due to the lack of $R, X$ measurement standards, $G$ is not a quantity directly measurable. Thus, $G$ can be obtained only analytically solving the quasi-stationary e.m. field configuration in the DR for simple geometries, or through electromagnetic numerical simulations. This is the major source of systematic (scale) effects.

The $u(Q_u)$ and $u(f_0)$ uncertainties give the major contributions to the final measurement uncertainties $u(Z)$ (apart from systematic effects). Their evaluation presents a complexity increased by the cryogenic (temperatures down to 4 K) and magnetic (fields up to 12 T) environment where measurements must be taken, so that the uncertainties $u(R)$ and $u(X)$ cannot be expected to be as small as in a completely controlled setup. $Q_u$ and $f_0$ are typically obtained from the frequency sweep measurement around $f_0$ of the two-port scattering $S$-parameters of the DR operated in transmission (time-domain measurements do not bring advantages due to high fluxon-motion losses in SCs). In less demanding environments, the standard '-3 dB method' is often employed: from the transmission $S$-parameters, the frequency of the resonance curve peak $S_M$ gives $f_0$ and the full width half maximum of the resonance curve $\Delta f_{0,-3\text{dB}}$ yields the loaded quality factor as $Q_l = f_0/\Delta f_{0,-3\text{dB}}$. In conventional measurements, the effect of the transmission lines is dealt with a proper calibration. In cryomagnetic cases this simple method does not bring sufficient accuracy and precision. First, the microphonic noise given by the vacuum pumping system, or by the flowing cryogenic gases, makes the $Q_u$ and $f_0$ measurement based on the '-3 dB method' highly scattered (much reduced precision). Second, the cross-coupling between the DR ports or the presence of higher order electromagnetic modes make the resonance curve asymmetric, and third, it is impossible to perform a full calibration of the whole transmission line at cryogenic temperatures (see later section "Measurement uncertainties"). Thus, unavoidable background contributions are added to the Lorentzian response to form the $S(f)$ measurements. For these reasons, different $f_0$ and $Q_u$ measurement strategies, based on fitting algorithms, are actively developed to overcome these issues [28]. In the case here presented, the transmission $S$-parameter, either $S_{12}$ or $S_{21}$, acquired with a Vector Network Analyzer (VNA) is fitted with the modified model given by:

$$S_{21}(f) = \left( \frac{S_M}{1 + i2Q_l \dfrac{f - f_0}{f_0}} + S_C \right) \cdot e^{i(\alpha + \beta f)}, \tag{6}$$

where $S_C$ is a frequency independent complex term used to take into account both cross-coupling between the ports of the DR and transmission line backgrounds, and where $\alpha$ and $\beta$ are real parameters used to include the frequency dependent phase delay. The fit is performed on the real and imaginary parts of $S_{21}$ using standard least squares curve fitting approaches based on the Levenberg-Marquardt algorithm on complex valued data and yields $f_0$ and the loaded quality factor $Q_l$. Finally, from the diameters of the $Q$-circles and lossy-circles traced by $S_{11}$ and $S_{22}$ on the complex plane, both the lossless and lossy components of the two coupling factors are determined to obtain the overall coupling factor $\beta$ through the transmission-mode $Q$-factor (TMQF) technique presented in [29], whence $Q_u = (1+\beta)Q_l$ [23]. We note that changes in $\beta$ need to be monitored, because of the thermal contraction/expansion of the metal enclosure, cables, loops.

In the next subsections, the performances of the measurement system and technique developed at Roma Tre University are detailed in terms of measurement sensitivity and uncertainty.

*Measurement sensitivity*

From Eq.s (4) and (5) the sensitivity functions are obtained:

$$c_1 = \frac{\partial Q_u}{\partial R} = -\frac{Q_u^2}{G}, \tag{7}$$

$$c_2 = \frac{\partial f_0}{\partial X} = -\frac{f_{0,ref}}{2G}. \tag{8}$$

Thus, to increase the measurement sensitivity, large $Q_u$ and small $G$ are required. Since $G = 4\pi f W/(\int_\Omega |H_{rf,\|}|^2 d\Omega)$ where $W$ is the total energy stored in the DR, $H_{rf,\|}$ the rf magnetic field component parallel to the surface of the sample and $\Omega$ the sample surface, small $G$ means large surface of the sample exposed to the $H_{rf,\|}$. Finally, $f_0$ is normally chosen to satisfy the need to characterize the sample at a specific frequency. As an example, for one of the resonators of the present work one has, at low $T$, $Q_u \sim 1.8 \times 10^4$, $f_0 \sim 15$ GHz and $G \sim 2.7 \times 10^3$ $\Omega$, whence $c_1 \sim -1.2 \times 10^5$ $\Omega^{-1}$ and $c_2 \sim 1.1 \times 10^7$ Hz $\Omega^{-1}$.

*Measurement uncertainties*

The uncertainties $u(R)$ and $u(X)$ are evaluated from Eq.s (4) and (5) with the standard uncertainty propagation procedure. These uncertainty sources are here discussed.

$u(Q_l)$ and $u(f_0)$ are obtained from the fitting procedure in a standard fashion from the numerically evaluated Jacobian matrix and the fit residuals. The relative standard uncertainties so evaluated are $\sim 0.02$ % for $Q_l$ and $\sim 5 \times 10^{-9}$ for $f_0$. Such low uncertainties require that the frequency span of $S_{21}(f)$ around $f_0$ is optimized. We found through experimental results and numerical simulations that $u(Q_l)$ and $u(f_0)$ cannot be minimized together. Best compromises lie in frequency span in the range 4-6 of the full width half-maximum of $|S_{21}(f)|$.

In cryogenic measurements, however, a major contribution to the uncertainties $u(Q_l)$ and $u(f_0)$ evaluation comes from the difficulty in performing the vector calibration of the whole transmission line, since thermal gradients are typically irreproducible and no microwave calibration standards are available for cryogenic $T$ and high $H$ fields operations. Actually, customized in-situ cryogenic calibration systems were designed [30-32]. However, these methods are not certified and traceable calibration systems (indeed, the only existing standard [24] about the measurement of SC $Z_s$ does not even address this point), they rely on components whose operation is certified only down to -50 °C, and no information are present on their behaviour in high magnetic fields. Moreover, they are based on bulky components and thus not suitable for the typical dimensions of cryostats for high magnetic fields: whence the need to deal with uncalibrated measurements (at least for the cryogenic portion of the line). Without a full calibration, the VNA cannot provide uncertainties on the measured $S$-parameters. To estimate $u(Q_u)$ and $u(f_0)$, we developed a room-temperature test system with similar $Q_l$ and operating frequencies, thus evaluating experimentally the contribution on uncertainties given by the lack of the calibration from the discrepancies of the $Q_l$ and $f_0$ measurements obtained applying (or not) the line calibration. We obtained experimental standard deviations of 2.2 % for $Q_l$ and $0.1 \times 10^{-6}$ for $f_0$ (in the $Q_l > 10^4$ range). These are then treated as uncertainties and combined with the previous ones to obtain the combined standard uncertainty on $u(Q_l)$ and $u(f_0)$.

The $u(\beta)$ was estimated for curves with SNR > 10 and $\beta > 0.05$, to be $\frac{u(\beta)}{\beta} < 3$ % with the implemented TMQF algorithm [29]. This uncertainty, combined with $u(Q_l)$, propagates to $u(Q_u)$. To reduce the contribution of $u(\beta)$ on $u(Q_u)$, the coupling of the resonator is set to the lowest possible value that yields a detectable signal (i.e. $\beta < 0.001$) in order to approximate $Q_l \approx Q_u$. The value of $\beta$ is monitored at each measuring temperature.

Finally, the $u(G)$ uncertainty is evaluated through Monte Carlo e.m. simulations of the DR, randomly varying all the dimensions and e.m. properties of the components assuming as gaussian all their distributions. The simulation results allow to assess $u(G)/G = 1$ %. All these uncertainties sources are then combined to obtain $u(Z)$.

## Experimental results

In this section sample measurements of the surface impedance on superconducting materials of largest technological interest for the advancement of research in fundamental experiments, such as Nb$_3$Sn ($T_c \approx$ 18 K) and YBa$_2$Cu$_3$O$_{7-x}$ ($T_c \approx$ 91 K), are shown and discussed, together with the extraction of the vortex motion parameters, $\rho_{ff}, f_c$ and $\chi$. Having to extract three observables, a minimum of three independent measurements is required. Alternatively, one can fit the data to Eq. (1) with a suitable number of desired quantities either neglected (typically, $\chi$=0) or taken as fit parameters [12, 13], but direct extraction from measurements is a more satisfying approach. This means that measurements of $Z = R + iX$ performed with a resonating technique at two different frequencies (at least) should be used, considering that Eq. (1) includes complex quantities.

As an example, we show measurements of the surface impedance on S1 sample. We developed a DR operating on two different transverse electric modes, the TE$_{011}$ at ∼ 16.4 GHz and TE$_{021}$ at ∼ 26.6 GHz. We use the analytical and numerical methods described above to obtain the loaded quality factor $Q_l$ and the resonance frequency $f_0$ as a function of the applied magnetic field at fixed temperature, as shown in Figure 3. In particular, we compare the results as obtained from the standard '-3 dB method' to the fitting of the resonance curves with the modified Lorentzian curve method described in the previous section. The precision of the '-3 dB method' is limited by the electronic noise and the finite frequency steps of the acquisition as well visible from the $\Delta f_0/f_{0,ref}$ measurement shown in Figure 3(a,c) results. Moreover, an evident systematic effect adds to the $Q_l$ measurement since the '-3 dB method' is affected by those non-idealities which are de-embedded from the $Q_l$ measurement by the fit procedure. The measurements shown in Figure 3 are obtained by setting the VNA output power at -7 dBm, 8001 acquisition points and 10 kHz of intermediate frequency (IF) bandwidth. The rf power is kept low enough to ensure to remain in the linear region of the investigated $Z$. For what concerns the choice of the number of points and the IF, there is a trade-off between the $S(f)$ curves acquisition rate, the needed points density near the resonance needed for the fitting procedure and the noise. Thus, these parameters are optimized depending on the characteristics of the measurement system and the measurement requirements in terms of noise and speed.

The sample $Z$ is then obtained through Eq.s (4) and (5). The next step is to determine the link between $Z$ and the sample $\rho$. In electromagnetically thick samples, $Z = \sqrt{i\omega\mu_0\rho}$; in thin SC films with thickness $t < \min(\delta, \lambda)$, with $\delta$ the skin penetration depth and $\lambda$ the London penetration depth, and when the SC film is deposited onto a dielectric substrate, $Z \simeq \rho/t$. We refer to [25] (and references therein) for a full discussion of the uncertainties introduced in such approximation. Since the variations $\Delta Z(H) = Z(H) - Z(H=0)$ induced by the magnetic field are measured, $\Delta Z(H)t \simeq \Delta\rho(H) \approx \rho_{vm}$. The first approximation relies on the film being electromagnetically thin, a geometrical property, the second relies on a physical condition, i.e. that the vortex motion resistivity is much larger than other superconducting contributions, which is realized not too close to the transition to the normal state. The $\rho_{vm}$ measurements are shown in Figure 3(b,d). From these, the vortex motion parameters are derived by means of Eq. (1) and shown in Figure 3(e,f,g).

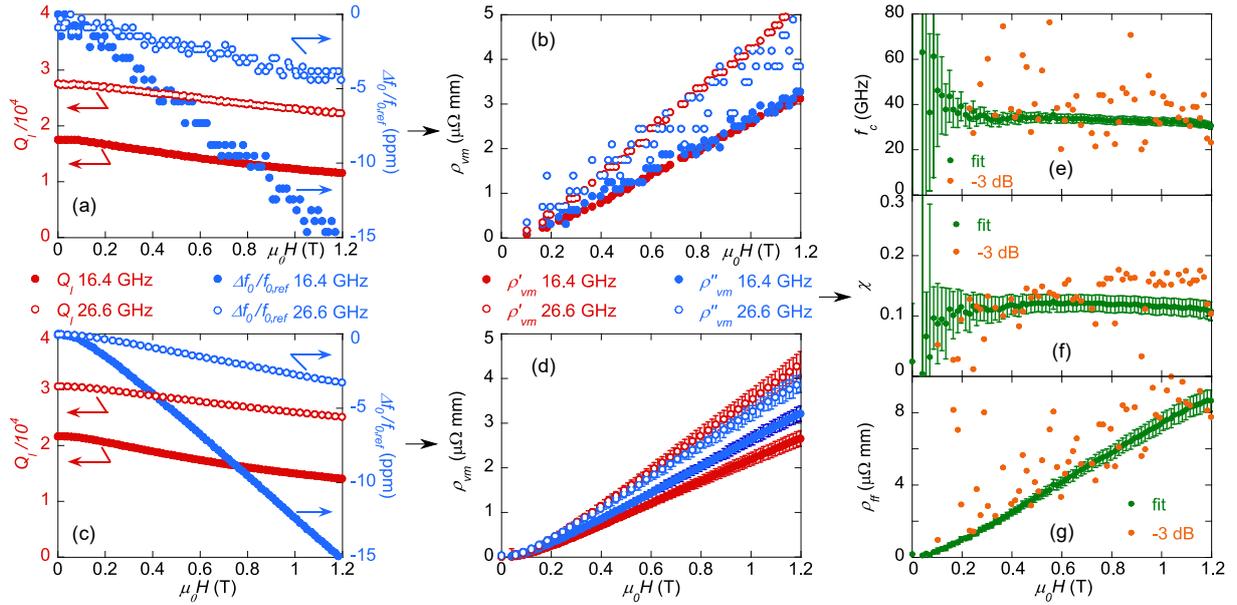

*Figure 3 – Panels (a,c): loaded quality factor $Q_l$ and relative variation of the resonance frequency $\Delta f_0/f_{0,ref}$ measured on a YBCO thin film (S1) at ~ 40 K and up to 1.2 T at 16.4 GHz (full symbols) and 26.6 GHz (empty symbols). Panels (b,d): vortex motion resistivity obtained from $Q_l$ and $\Delta f_0/f_{0,ref}$ measurements. Upper panels (a,b): data obtained with the '-3dB method' (no uncertainty bar reported); lower panels (c,d): data obtained through the fitting procedure (when not visible, uncertainty bars are contained in the size of the markers). Panels (e,f,g): comparison of the characteristic frequency $f_c$, thermal creep factor $\chi$ and flux-flow resistivity $\rho_{ff}$ obtained from the vortex motion resistivity as determined with the '-3 dB method' (orange symbols) and from the fitting procedure (green symbols).*

It can be noted how measurements of the surface reactance require a rather accurate fitting of $S_{21}$ and $S_{12}$. The '-3 dB method' yields unacceptably large uncertainties in the vortex parameters (see the scattering of the data in Figures 3(e,f,g)).

The field dependence of the obtained parameters yields information on pinning that is very useful for the optimization of the SC material. In this specific case, for example, the field independence of $f_c$ and $\chi$ identifies, within the fluxon dynamics theory, a regime in which the density of the pinning centres in the materials is higher than the density of the fluxons, so that further optimization is possible only by fine tuning the kind of defects and not the density.

A comparison between S2 (YBCO) and S3 (Nb$_3$Sn), is very interesting. The surface impedance of a bulk sample of Nb$_3$Sn was measured with a single tone DR at ~ 14.9 GHz, down to 4 K and up to 12 T [26] The low operating temperature allows to safely neglect χ, so that single-frequency $\rho_{vm} = \rho'_{vm} + i\rho''_{vm}$ still yield $f_c$ and $\rho_{ff}$. The results are shown in Figure 4 up to 12 T, at 6 K for Nb$_3$Sn and 27 K for YBCO. Due to the different critical temperatures, $T_c$ ~ 18.0 K in Nb$_3$Sn and $T_c$ ~ 91.0 K in YBCO, the selected temperatures correspond to similar reduced temperature $T/T_c \approx 0.3$.

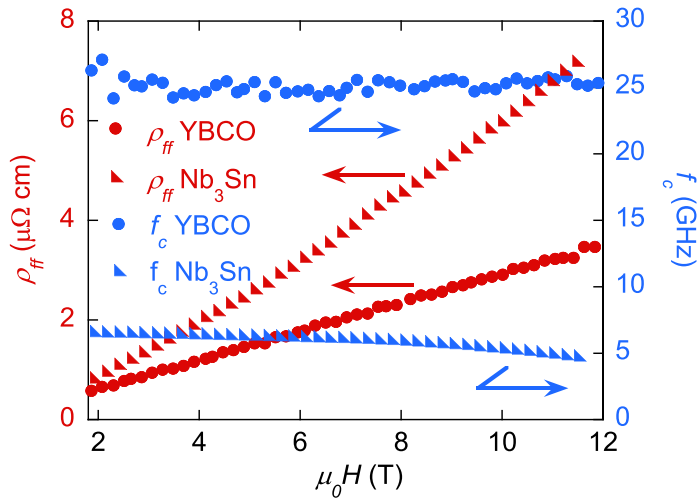

*Figure 4 – Flux-flow resistivity (on the left axis in red) and the characteristic frequency (on the right axis in blue) measured at similar reduced temperature $T/T_c$ on a bulk Nb$_3$Sn sample at 6 K (triangles) and on a YBCO thin film (full dots) at 27 K, for fields up to 12 T.*

For practical use of SCs at high frequency in large dc magnetic fields, operation at $f \ll f_c$ is required. $f_c$ in Nb$_3$Sn at e.g. 10 T is ~ 5 times smaller than that of YBCO, which is rather discouraging for rf applications. However, the field dependence of $f_c$ points to a different pinning regime, known as collective pinning, in which a low density of pinning centres is present in the SC with respect to the density of the fluxons. This is not the most effective pinning regime, thus further material improvements can be made, e.g. by adding artificial pinning centres in the SC to optimize the high frequency behaviour of Nb$_3$Sn in the mixed state. This is a completely new paradigm in high-frequency applications of SCs: whereas for accelerator cavities (operating in zero field) one looks for the purest SC material, a different approach may be required for the requirements of new experiments. Additionally, $\rho_{ff}(H)$ has a similar slope in both SCs. This means that in the high frequency limit $f \gg f_c$ the losses in Nb$_3$Sn at 6 K are of the same order than those in YBCO at 27 K in the same field range up to 12 T, with only a factor of ~2 between the two. Since $\rho_{ff}$ is directly related to the microscopic properties of the material, tuning of this property by material science is not expected.

The present measurements thus indicate that by engineering the PCs it can be possible to improve the high-frequency, in-field properties of a "SC workhorse" like Nb$_3$Sn, while the high $f_c$ of YBCO makes this material already mature, from the point of view of the physical properties, for the perspective high frequency applications of SCs in the mixed state. Indeed, the properties of various commercial YBCO tapes have been recently studied at 8 GHz and up to 9 T [13], showing that even dc-optimized, industrial YBCO tapes could fulfil the requirements of the aforementioned large scales applications. However, it must be noticed that commercial tapes are optimized for dc operations through the introduction of such a large density of PCs that they have detrimental effects on $\rho_{ff}$: despite the large $f_c$ found in some nanoengineered YBCO tapes, the slope of $\rho_{ff}(H)$, on the same tapes, was measured to be ~ 3 times larger than that obtained in pristine YBCO [13]. We point out (see Figure 1) that both $f_c$ and $\rho_{ff}$ are instrumental in assessing the suitability of SCs for these kinds of applications. Figure 5 shows a summary of such framework for several potentially interesting superconductors including the results obtained on S1, S2, S3, S4 and S5 samples. As an example, FeSe$_{0.5}$Te$_{0.5}$ (S4), belonging to the new family of the iron based SCs still unexplored for high-frequency applications, exhibits a large $f_c$ but also a particularly large $\rho_{ff}$ due to its complex electronic states structure, while MgB$_2$ (S5) is only slightly better than Nb$_3$Sn. How $\rho_{ff}$ can be tuned (in particular on FeSe$_{0.5}$Te$_{0.5}$) is still a matter of fundamental research. As it can be seen from Fig. 5, a reliable determination of both $f_c$ and $\rho_{ff}$ is essential. From these observations and from the experimental results, it emerges that none of the SC candidates for rf applications in a high dc magnetic field are fully optimized yet: by limiting the discussion to Nb$_3$Sn and YBCO, while Nb$_3$Sn would benefit from a higher density of PCs, YBCO requires PCs that affect less the microscopic scattering mechanisms.

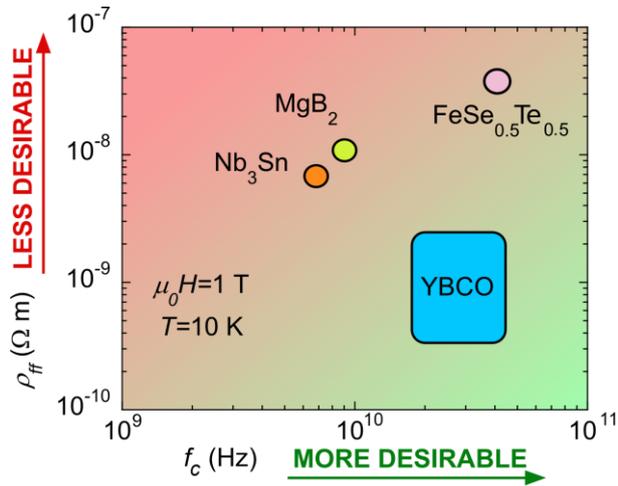

Figure 5 – Plot of the flux-flow resistivity $\rho_{ff}$ and characteristic frequency $f_c$ measured at 10 K and 1 T on different SCs: $Nb_3Sn$ (orange), $MgB_2$ (green), YBCO (light blue) and $FeSe_{0.5}Te_{0.5}$ (pink).

## Conclusions

The surface impedance $Z$ measurement of superconductors (SCs) in the mixed state is a fundamental step in the test and development of materials with always increased performances.

This type of measurement provides useful information both in view of rf technological applications, such as the aforementioned beam screens for particle accelerators or the realization of haloscopes for dark matter particles detection experiments, and for scientific aspects related to the study of the physics of SCs. Given the emerging technological interest, a metrological study of the performance of these techniques is essential to obtain reliable measurements. In this framework, multifrequency dielectric resonators sided with suitable data analysis procedures are powerful tools for a correct evaluation of the vortex motion parameters in SCs. A comparison of the parameters as measured in different materials allows to identify the margins for improvement which can be achieved through the engineering of the materials.

It should be mentioned that, in particular for particle accelerators beam screens, studies such as the present one are preliminary since only the linear response is addressed. Measurements of $Z$ at frequencies closer to operation (~ 1 GHz) and, more importantly, at higher rf power will be needed to assess the feasibility of superconducting beam screens.

In this framework, the need to perform accurate and precise $Z$ measurements at cryogenic temperatures and high frequency, high dc fields and large rf powers, is a metrological challenging journey that has only begun.

## Acknowledgements

This work has been partially carried out within the framework of the EUROfusion Consortium and has received funding from the Euratom Research and Training Programme 2019-2020 under grant agreement No 633053. The views and opinions expressed herein do not necessarily reflect those of the European Commission. The work has been also partially supported by MIUR-PRIN project "HIBiSCUS" - grant no. 201785KWLE. The authors warmly thank T. Spina, from the "Fermi National Accelerator Laboratory" (USA), and R. Flükiger, from the University of Geneva (Switzerland), for having provided the $Nb_3Sn$ sample; G. Celentano, V. Pinto, A.A. Armenio, from the "Agenzia nazionale per le nuove tecnologie, l'energia e lo sviluppo economico sostenibile" - ENEA (Italy), for the YBCO samples; V. Braccini, from "Consiglio Nazionale delle Ricerche - SuPerconducting and other INnovative materials and devices institute" - CNR-SPIN (Italy), for the $FeSe_{0.5}Te_{0.5}$ samples and A. Crisan, from the "National Institute of Materials Physics" - NIMP (Romania), for the $MgB_2$ samples.

# References

[1] H. Padamsee, "The science and technology of superconducting cavities for accelerators", *Supercond. Sci. Technol.*, vol. 14, no. 4, R28, 2001.

[2] Anlage, S. M., "Microwave Superconductivity", *IEEE J. Microw.*, vol. 1, no. 1, pp. 389-402, 2021.

[3] S. Calatroni, "HTS coatings for impedance reduction in particle accelerators: case study for the FCC at CERN". *IEEE Trans. Appl. Supercond.*, vol. 26, no. 3, art. id 3500204, 2016.

[4] P. Gan et al., "Design study of an YBCO-coated beam screen for the super proton-proton collider bending magnets", *Rev. Sci. Instrum.*, vol. 89, no. 4, art. id 045114, 2018.

[5] D. Alesini et al., "Galactic axions search with a superconducting resonant cavity", *Phys. Rev. D*, vol. 99, no. 10, art. id 101101, 2019.

[6] A. Grudjev, "First thoughts on required RF testing infrastructure," *in Proc. Muon Collider Workshop,* CERN Geneva, 2019. [Online]. Available: https://indico.cern.ch/event/845054/

[7] S. R. Foltyn et al., "Materials science challenges for high-temperature superconducting wire", *Nat. Mater.*, vol. 6, no. 9, pp. 631-642, 2007.

[8] J. I. Gittleman and B. Rosenblum, "Radio-frequency resistance in the mixed state for subcritical currents", *Phys. Rev. Lett.*, vol. 16, no. 17, pp. 734-736, 1966.

[9] M. Tinkham, *Introduction to superconductivity*, Courier Corporation, 2004.

[10] M. Golosovsky, M. Tsindlekht and D. Davidov, "High-frequency vortex dynamics in YBa$_2$Cu$_3$O$_7$", *Supercond. Sci. Technol.*, vol. 9, no. 1, pp. 1-15, 1996.

[11] M. W. Coffey and J. R. Clem, "Unified theory of effects of vortex pinning and flux creep upon the rf surface impedance of type-II superconductors", *Phys. Rev. Lett.*, vol. 67, no. 3, pp. 386-389, 1991.

[12] N. Pompeo, and E. Silva, "Reliable determination of vortex parameters from measurements of the microwave complex resistivity", *Phys. Rev. B*, vol. 78, no. 9, art. id 094503, 2008.

[13] A. Romanov et al., "High frequency response of thick REBCO coated conductors in the framework of the FCC study", *Sci. Rep.*, vol. 10, no. 1, art. id 12325, 2020.

[14] D. H. Wu, J. C. Booth and S. M. Anlage, S. M., "Frequency and field variation of vortex dynamics in YBa$_2$Cu$_3$O$_{7-\delta}$". *Phys. Rev. Lett.*, vol. 75, no. 3, pp. 525-528, 1995.

[15] E. Silva et al., "Wideband surface impedance measurements in superconducting films", *IEEE Trans. Instrum. Meas.*, vol. 65, no. 5, pp. 1120-1129, 2016.

[16] V. Pinto et al., "Chemical Solution Deposition of YBCO Films with Gd Excess", *Coatings*, vol. 10, no. 9, art. id. 860, 2020.

[17] A. Angrisani Armenio et al., "Analysis of the growth process and pinning mechanism of low-fluorine MOD YBa$_2$Cu$_3$O$_{7-\delta}$ films with and without BaZrO$_3$ artificial pinning centers", *IEEE Trans. Appl. Supercond.*, vol. 25, no. 3, art. id. 6605205, 2015.

[18] T. Spina, "Proton irradiation effects on Nb$_3$Sn wires and thin platelets in view of High Luminosity LHC upgrade", *PhD Thesis*, University of Geneva, 2015.

[19] V. Braccini et al., "Highly effective and isotropic pinning in epitaxial Fe(Se,Te) thin films grown on CaF$_2$ substrates" *Appl. Phys. Lett.*, vol. 103, no. 17, art. id. 172601, 2013.

[20] V. Sandu et al., "Tellurium addition as a solution to improve compactness of ex-situ processed MgB2-SiC superconducting tapes", *Supercond. Sci. Technol.*, vol. 29, no. 6, art. id. 065012, 2016.

[21] J. C. Booth, D. H. Wu, and S. M. Anlage, "A broadband method for the measurement of the surface impedance of thin films at microwave frequencies", *Rev. Sci. Instrum.*, vol. 65, no. 6, pp. 2082-2090, 1994.

[22] H. Kitano, et al., "Broadband method for precise microwave spectroscopy of superconducting thin films near the critical temperature", *Rev. Sci. Instrum.*, vol. 79, no. 7, art. id. 074701, 2008.

[23] L. F. Chen et al., *"Microwave electronics: measurement and materials characterization"*, John Wiley & Sons, 2004.

[24] International Electrotechnical Commission IEC 61788-7:2020, Superconductivity - Part 7: Electronic characteristic measurements - Surface resistance of high-temperature superconductors at microwave frequencies.

[25] A. Alimenti et al., "Challenging microwave resonant measurement techniques for conducting material characterization", *Meas. Sci. Technol.*, vol. 30, no. 6, art. id 065601, 2019.

[26] A. Alimenti et al., "Microwave measurements of the high magnetic field vortex motion pinning parameters in $Nb_3Sn$", *Supercond. Sci. Technol.*, vol. 34, no. 1, art. id 014003, 2020.

[27] J. Krupka, et al., "Microwave complex conductivity of the YBCO thin films as a function of static external magnetic field", *Appl. Phys. Lett.*, vol. 104, no. 10, art. id. 102603, 2014.

[28] P. J. Petersan and S. M. Anlage, "Measurement of resonant frequency and quality factor of microwave resonators: Comparison of methods", *J. Appl. Phys.*, vol. 84, no. 6, pp. 3392-3402, 1998.

[29] K. Leong and J. Mazierska, "Precise measurements of the Q factor of dielectric resonators in the transmission mode-accounting for noise, crosstalk, delay of uncalibrated lines, coupling loss, and coupling reactance", *IEEE Trans. Microw. Theory*, vol. 50, no. 9, pp. 2115-2127, 2002.

[30] J. H. Yeh and S. M. Anlage, "In situ broadband cryogenic calibration for two-port superconducting microwave resonators", *Rev. Sci. Instrum.*, vol. 84, no. 3, art. id 034706, 2013.

[31] G. Crupi et al., "Measurement-Based Extraction and Analysis of a Temperature-Dependent Equivalent-Circuit Model for a SAW Resonator: From Room Down to Cryogenic Temperatures", *IEEE Sens. J.*, vol. 21, no. 10, 12202-12211, 2021.

[32] D. E. Oates, R. L. Slattery, and D. J. Hover, "Cryogenic test fixture for two-port calibration at 4.2 K and above," in 2017 89th ARFTG Microwave Measurement Conference (ARFTG), 2017, pp. 1–4.

## Short bios

*Andrea Alimenti* (andrea.alimenti@uniroma3.it) is a Postdoctoral Researcher in the Department of Engineering at Roma Tre University in Roma, Italy. From the same university he received his B. Sci. and M. Sci in Electronic Engineering in 2014 and 2017, respectively, and he got the Ph.D. in Applied Electronics in 2021. His main research interest is the development of microwave measurement techniques for the characterization and study of dielectrics, conductors and superconductors.

*Nicola Pompeo* (nicola.pompeo@uniroma3.it) is Assistant Professor in Physics in the Department of Engineering at the Roma Tre University, Italy. He got his M. Sci. in Electronic Engineering in 1998 and, after some years as system engineer in the industry, he got a Ph. D. in Physics in 2006. His research deals mainly with the theory and the experimental study of the high frequency electrodynamics of superconductors, including the design, realization and optimization of systems for the surface impedance measurement.

*Kostiantyn Torokhtii* (kostiantyn.torokhtii@uniroma3.it) is a Technical Researcher in the Department of Engineering at the Roma Tre University, Italy. He received M.Sci. degree in Cryogenic Technique and Technology in 2007 from NTU "KhPI", Kharkiv, Ukraine, and a Ph.D. degree in Biomedical Electronics, Electromagnetism and Telecommunication in 2013 from Roma Tre University, Rome, Italy. His research interests include microwave measurement techniques and their applications, surface impedance measurements, the study of complex superconducting structures and unconventional superconductors.

*Enrico Silva* (enrico.silva@uniroma3.it) is Full Professor of Electrical and Electronic Measurements in the Department of Engineering at Roma Tre University in Rome, Italy. He received his M. Sci. in Physics in 1990 and Ph. D. in Applied Electromagnetism in 1994 from University "La Sapienza", Rome. Enrico Silva is member of the IEEE Council on Superconductivity AdCom. The research interests of Enrico Silva include the development of novel or improved microwave methods in difficult environment, such as the study of superconductors.